\documentclass[reprint,superscriptaddress,showpacs,twocolumn,amsmath,amssymb,aps,prl]{revtex4-1}

\usepackage{graphicx}
\usepackage{dcolumn}
\usepackage{bm}
\usepackage[dvipsnames]{xcolor}
\usepackage[english]{babel}
\usepackage{tabularx}
\usepackage{booktabs}

\begin{document}

\title{Unconventional quantum oscillations in mesoscopic rings of spin-triplet superconductor Sr$_2$RuO$_4$}

\author{X.~Cai}
\affiliation{Department of Physics and Materials Research Institute, The Pennsylvania State University, University Park, Pennsylvania 16802, USA}

\author{Y.~A.~Ying}
\affiliation{Department of Physics and Materials Research Institute, The Pennsylvania State University, University Park, Pennsylvania 16802, USA}

\author{N.~E.~Staley}
\altaffiliation{Current Address: Department of Physics, Massachusetts Institute of Technology, Cambridge, MA 02139}
\affiliation{Department of Physics and Materials Research Institute, The Pennsylvania State University, University Park, Pennsylvania 16802, USA}

\author{Y.~Xin}
\affiliation{National High Magnetic Field Laboratory, Florida State University, Tallahassee, Florida 32310, USA}

\author{D.~Fobes}
\affiliation{Department of Physics, Tulane University, New Orleans, Louisiana 70118, USA}

\author{T.~J.~Liu}
\affiliation{Department of Physics, Tulane University, New Orleans, Louisiana 70118, USA}

\author{Z.~Q.~Mao}
\affiliation{Department of Physics, Tulane University, New Orleans, Louisiana 70118, USA}

\author{Y.~Liu}
\email{liu@phys.psu.edu}
\affiliation{Department of Physics and Materials Research Institute, The Pennsylvania State University, University Park, Pennsylvania 16802, USA}

\date{\today}

\begin{abstract}
Odd-parity, spin-triplet superconductor Sr$_2$RuO$_4$ has been found to feature exotic vortex physics including half-flux quanta trapped in a doubly connected sample and the formation of vortex lattices at low fields. The consequences of these vortex states on the low-temperature magnetoresistive behavior of mesoscopic samples of Sr$_2$RuO$_4$ were investigated in this work using ring device fabricated on mechanically exfoliated single crystals of Sr$_2$RuO$_4$ by photolithography and focused ion beam. With the magnetic field applied perpendicular to the in-plane direction, thin-wall rings of Sr$_2$RuO$_4$ were found to exhibit pronounced quantum oscillations with a conventional period of the full-flux quantum even though the unexpectedly large amplitude and the number of oscillations suggest the observation of vortex-flow-dominated magnetoresistance oscillations rather than a conventional Little-Parks effect. For rings with a thick wall, two distinct periods of quantum oscillations were found in high and low field regimes, respectively, which we argue to be associated with the ``lock-in" of a vortex lattice in these thick-wall rings.  No evidence for half-flux-quantum resistance oscillations were identified in any sample measured so far without the presence of an in-plane field. 

\end{abstract}

\maketitle

Sr$_2$RuO$_4$, the only superconducting layered perovskite without Cu, features an odd-parity, spin-triplet, likely chiral $p$-wave pairing state~\cite{Mackenzie2003,Nelson2004,Xia2006,Kidwingira2006,Maeno2012}. In addition to an exotic pairing symmetry, unconventional flux states have also been observed in this superconductor. Neutron scattering study revealed the existence of a square rather than the traditional triangular vortex lattice over a large range of temperatures and magnetic fields applied along the $c$ axis~\cite{Riseman1998}. Surprisingly, vortex lattices were observed~\cite{Curran2011} at a field much lower than the lower critical field $H_{c1\parallel c}$ = 50 Oe \cite{Akima1999}. More specifically, the triangular Abrikosov vortex lattice was observed at a field as low as 5.4 Oe and a well-ordered square lattice was found to form for fields above 12.7 Oe. Even though the precise nature of the transition between the two types of vortex lattices was identified, the existence of the vortex lattices at such low fields reflects ease of the vortex lattice formation and the dominance of vortex-vortex interaction over pinning potential energies. For small magnetic fields applied away from the $c$ axis, vortex coalescence overcoming the vortex-vortex repulsive interaction was observed~\cite{Dolocan2005}. Most recently cantilever magnetometry measurements revealed half-height step features in the magnetization with a finite field applied in the in-plane direction, suggesting the existence of a $\Phi_0$/2 flux state~\cite{Kee2000} ($\Phi_0$= $h$/2$e$, where $h$ is the Planck constant and $e$ the elemental charge) in micron-sized doubly connected samples of Sr$_2$RuO$_4$~\cite{Jang2011}. The existence of a $\Phi_0$/2 vortex, yet to be observed directly,  is particularly important for the pursuit of fault-tolerant topological quantum computing based on non-Abelian Majorana fermions~\cite{Nayak2008} as it was proposed that a $\Phi_0$/2 vortex would carry a Majorana mode in its normal core, which is the basis for the topological quantum computing~\cite{DasSarma2006}.

To explore the consequences of these exotic flux states on the low-temperature magnetoresistive behavior of Sr$_2$RuO$_4$, low-temperature measurements on mesoscopic samples of this superconductor are highly desired. Magnetoresistance oscillation measurements, which require the fabrication of small superconducting structures, are capable of providing direct observation of the flux quantization in superconductors. The existence of a macroscopic quantum mechanical wave function demands the fluxoid enclosed in a doubly connected superconductor be quantized. Consequently, the superfluid velocity ($v_s$) is a periodic function of the applied magnetic flux with a periodicity equal to $\Phi_0$. Such a periodicity in $v_s$ results in an oscillating superconducting transition temperature ($T_c$) and hence the sample resistance in the transition regime, known as the Little-Parks (LP) effect. The experimental study of the LP effect has so far been focused on conventional $s$-wave superconductors because of either the difficulty in sample fabrication, or the commonly found short coherence length ($\xi$) for unconventional superconducting materials such as high-$T_c$ superconductors with $d$-wave pairing. A recent attempt on nanopatterned high-$T_c$ superconducting films were found to exhibit magnetoresistance oscillations with a large amplitude unexpected from the LP effect~\cite{Sochnikov2010}. It was suggested that in a similar structure of Sr$_2$RuO$_4$, $\Phi_0$/2 states can be well distinguished from those of $\Phi_0$ by transport measurements~\cite{Vakaryuk2011}. 

In this Letter, we present magnetoresistance measurements on mesoscopic superconducting rings of Sr$_2$RuO$_4$.  The preparation of such samples would typically require thin films suitable for nanofabrications. However, superconducting films of Sr$_2$RuO$_4$ are extremely difficult to synthesize. Even though the preparation of a superconducting film of Sr$_2$RuO$_4$ was reported recently~\cite{Krockenberger2010}, no superconducting films have been grown since then~\cite{Unpublished}. In addition, either patterning a single loop that is both small and superconducting or making electrical contact to such a loop is a significant nanofabrication challenge given the sensitivity of superconductivity in Sr$_2$RuO$_4$ to disorder. We were able to meet these challenges using thin, flat crystals prepared by mechanical cleaving from bulk single crystals. 

Device fabrication in this work starts with the preparation of small crystals of Sr$_2$RuO$_4$ with a typical dimension of 20 to 30 $\mu$m in lateral size and 0.2 to 0.6 $\mu$m in thickness. Bulk single crystals of Sr$_2$RuO$_4$ were grown by a floating zone method. To minimize the formation of Ru microdomains in the crystals and therefore promote easy cleaving along the in-plane direction, the Ru over compensation was reduced for the growth of most crystals used in this work. By crushing a freshly cleaved single crystal onto a flat substrate, large numbers of small Sr$_2$RuO$_4$ crystals were obtained. Photolithography was used to prepare 4 or 6-point, Ti/Au electrical contacts to the thin, flat crystals. After a 200 nm thick SiO$_2$ protective layer was deposited on both the contacts and the crystals, micron-sized rings with 4 leads were patterned at the center of the crystals using a focused ion beam (FIB) of 30 keV Ga ions with a beam current of 50 pA (corresponding to a beam size of 9.5 nm). The Sr$_2$RuO$_4$ crystal near large Ti/Au contacts was also cut to separate the leads enabling the transport measurements on the rings [Fig.~1(a)]. The resistance oscillations of the rings were measured in a dilution refrigerator with a base temperature of 20 mK, using a dc technique. 

\begin{table}
\caption{Sample dimensions. $r_m$ is the mid-point radius, $h$ the height of the ring, and $w$ the wall thickness measured on the top and at the bottom of the ring, receptively. Measurement uncertainty: $\pm$ 10 nm.}
\begin{tabular*}{0.45\textwidth}{@{\extracolsep{\fill}}l*{5}{c}r}
\hline\hline
             & $r_m$ (nm) & $h$ ($n$m) & \multicolumn{2}{c}{$w$ (nm)}\\
              \cline{4-5}
             & & & top & bottom \\
\hline
Sample 1      & 480 & 450 & 250 & 390 \\
Sample 2      & 440 & 400 & 160 & 230 \\
Sample 3      & 460 & 520 & 250 & 470 \\
\hline\hline
\end{tabular*}
\end{table}

To determine the wall thickness of the superconducting part of the nanofabricated rings of Sr$_2$RuO$_4$ needed for estimating the resistance oscillation period, analytic tools were used to characterize these rings and measure the thickness of the damaged layer caused by the fabrication process. In particular, scanning transmission electron microscopy (STEM) images of a ring cross-section cut by FIB revealed damaged regions near the sidewall, extending about 20 nm deep into the wall of the Sr$_2$RuO$_4$ ring [Figs.~1(c) and 1(d)]. The damaged layer features Sr, Ru, and Ga species resulting from Ga implantation as well as material re-deposition during FIB patterning. However, as shown in Fig.~1(b), the interior of the Sr$_2$RuO$_4$ ring maintains excellent crystallinity. Rings of Sr$_2$RuO$_4$ so fabricated, even with a wall thickness down to 200 nm, were found to be superconducting [Figs.~2(a) and 2(b)]. Experimental results on three rings with the sample dimensions detailed in Table I, are presented below.

\begin{figure}[t!]
\includegraphics[width=3.1in]{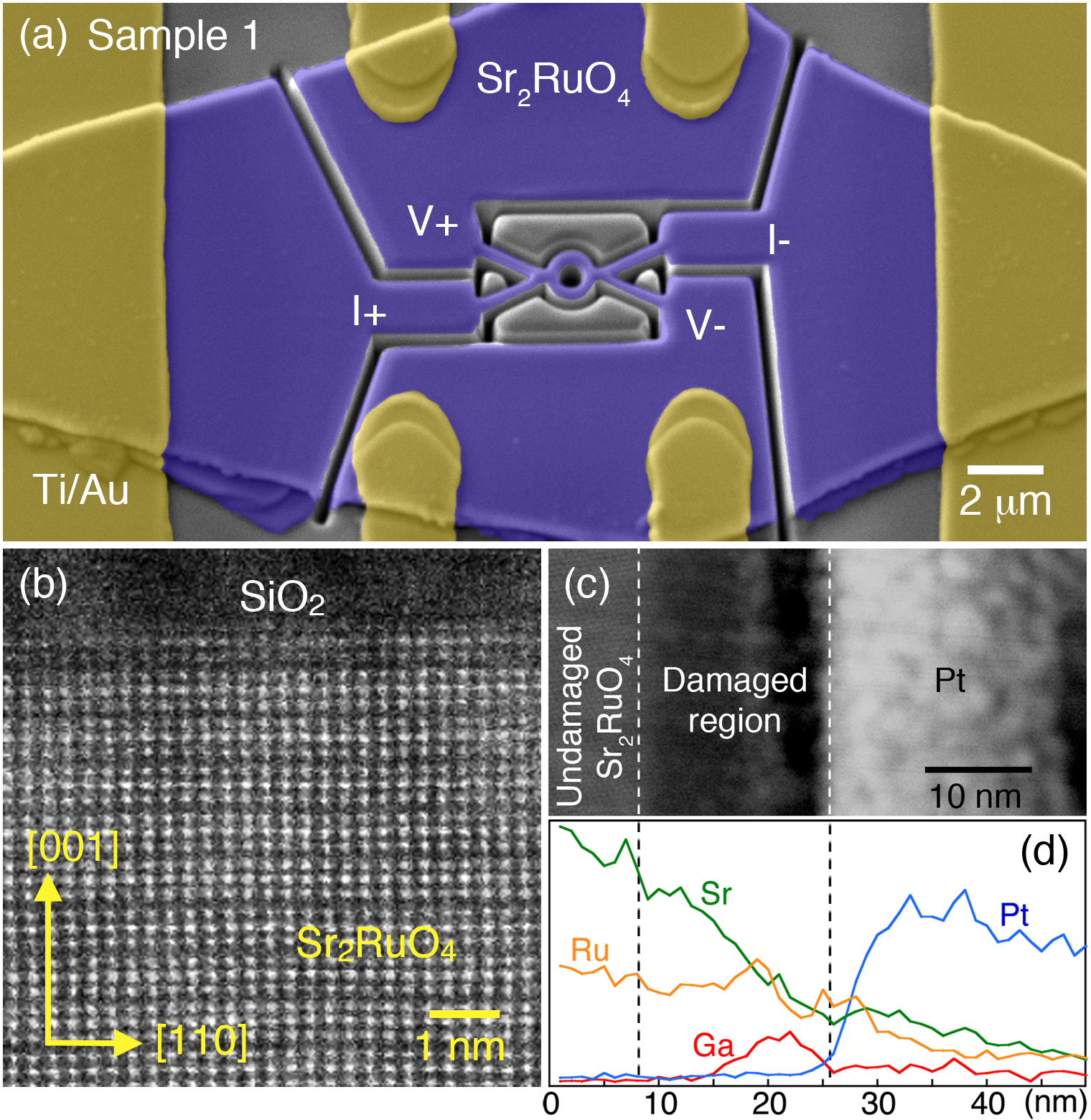} 
\caption{\label{fig:epsart} (Color online) (a) False-color SEM image of Sample 1, showing the Sr$_2$RuO$_4$ crystal, the Ti/Au leads (yellow), and the ring device (blue). (b) Atomic-resolution STEM high-angle annular dark field (HAADF) image showing that the crystalline structure of Sr$_2$RuO$_4$ extended to the top layer of the crystal. Brighter dots are Ru atoms. (c) STEM image taken on the ring sidewall. Pt was deposited as a protection layer. A 0.11 nm electron probe was moved from left to right across the boundary collecting the energy dispersive X-ray spectra (EDS) elemental line profiles across the interface. (d) Sr, Ru, Ga and Pt atomic composition
profiles obtained from EDS. The probe position is matched with the position in (c) as marked by the dashed lines.}
\end{figure}

\begin{figure}[t!]
\includegraphics[width=3.1in]{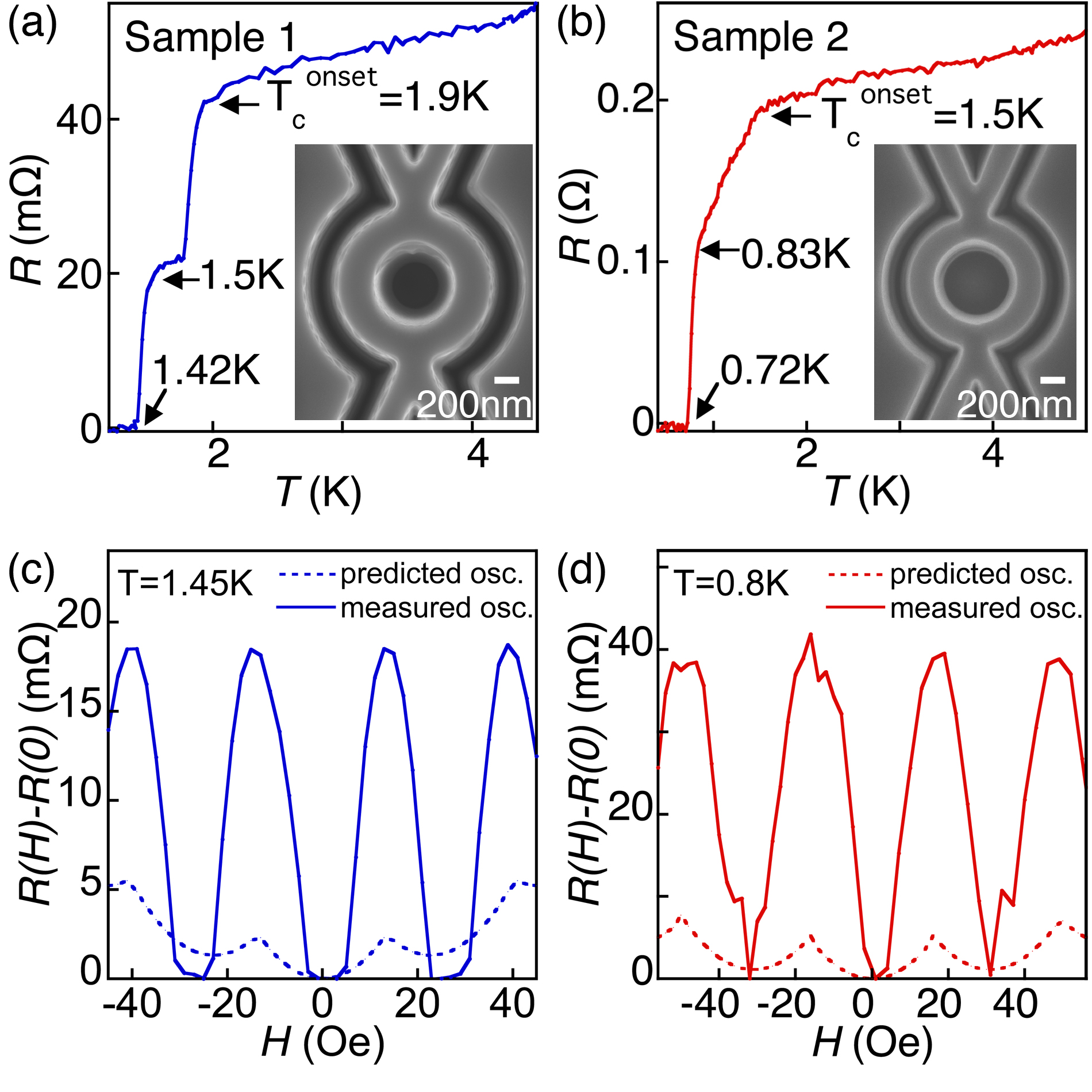}
\caption{\label{fig:epsart} (Color online) Temperature dependence of sample resistance $R(T)$ for (a) Sample 1 and (b) Sample 2. Insets: SEM images of the ring device. (c) Measured magnetoresistance oscillations of Sample 1 (solid curves). The predicted amplitude of resistance oscillations for the Little-Parks effect at low fields (dashed curves). For this estimate, $dR/dT$ = 0.245 $\Omega$/K was used, corresponding to the measured slope in $R(T)$ at 1.45 K. (d) Measured and predicted oscillations of Sample 2 at 0.8 K. $dR/dT$ = 0.567 $\Omega$/K was used.}
\end{figure}

Applying a magnetic field perpendicular to the plane of the ring ($H\parallel c$), we found pronounced magnetoresistance oscillations in the transition region. Based on the Ginzburg-Landau theory of the LP oscillations observed in a conventional $s$-wave superconductor loop with a wall thickness $w$, the $T_c$ oscillations are given by \cite{Groff1968}
\begin{equation}
 \frac{\Delta
T_c}{T_c}=\frac{\xi(0)^2}{r^2_m}\left[\left(n'-\frac{\Phi}{\Phi_0}\right)^2(1+a^2)+\frac{4}{3}n'^2a^2\right],
\end{equation}
where $\xi(0)$ is the zero-temperature coherence length, $r_m = (r_{outer} +r_{inner})/2$, $\Phi=\pi r^2_m H$, $a = w/2r_m$, $H$ the applied field, and $n'=\frac{n}{1+a^2}$. Here $n$ is an integer that takes a suitable value to maximize $T_c(H)$ as the field is ramped, leading to a  $T_c$ oscillation. The field increment between two successive maxima in $T_c$ for the $\Phi_0$ state is $\Delta H=\Phi_0/[\pi r_m^2(1+a^2)]$. The amplitude of the resistance oscillations can be estimated using $\Delta R=\Delta T_c (dR/dT)$, where $dR/dT$ is the slope of the $R(T)$ curve in the transition region. Using the in-plane zero-temperature coherence length value of Sr$_2$RuO$_4$, $\xi_{ab}(0)$ = 66 nm, with $T_c$ = 1.5 K, and the dimensions of the rings, as a rough estimate, the calculated LP magnetoresistance oscillations are shown as the dashed curves in Figs. 2(c) and 2(d). The observed amplitude of the resistance oscillations [solid curves in Figs. 2(c) and 2(d)] is an order of magnitude larger than the predicted. 

Furthermore, a large number of pronounced oscillations were observed at temperatures far below $T_c$. In a ring (Sample 2, Inset in Fig.~2(b)), the observed amplitude of the resistance oscillation, $\Delta R$, is essentially as large as $R_N$, the normal state resistance, at all temperatures up to 0.7 K [Fig.~3a]. For the rings where $w$ is comparable to $\xi(0)$ ($w\sim 3\xi(0)$ for Sample 2), the enhanced magnetoresistance oscillations can be explained in a model based on voltages induced by vortices moving in and out of a superconducting ring \cite{Vakaryuk2011}. The motion of vortices across the wall is driven by the combined effect of the measurement current and the circulating current $j_s$ demanded by fluxoid quantization. Consequently, as the barrier potential for vortex moving in and out the ring depends on the applied flux periodically, the vortex flow rate is a periodic function of the applied flux. This leads to an oscillation in the voltage (causing a sample resistance) across the sample. Compared with those observed in high-$T_c$ superconductors \cite{Sochnikov2010}, the resistance oscillations in Sr$_2$RuO$_4$ rings are more pronounced. The observed oscillation period, $\Delta H \approx$ 31.4 Oe (as indicated by the dashed lines in Fig.~3(a)), is consistent with that of $\Phi_0$. According to Eq.~(1), $\Phi_0$ corresponds to $32\pm2$ Oe, where the error bar is estimated from the variation of the wall-thickness, as well as the measurement uncertainty in sample dimensions.

\begin{figure}[t!]
\includegraphics[width=3.34in]{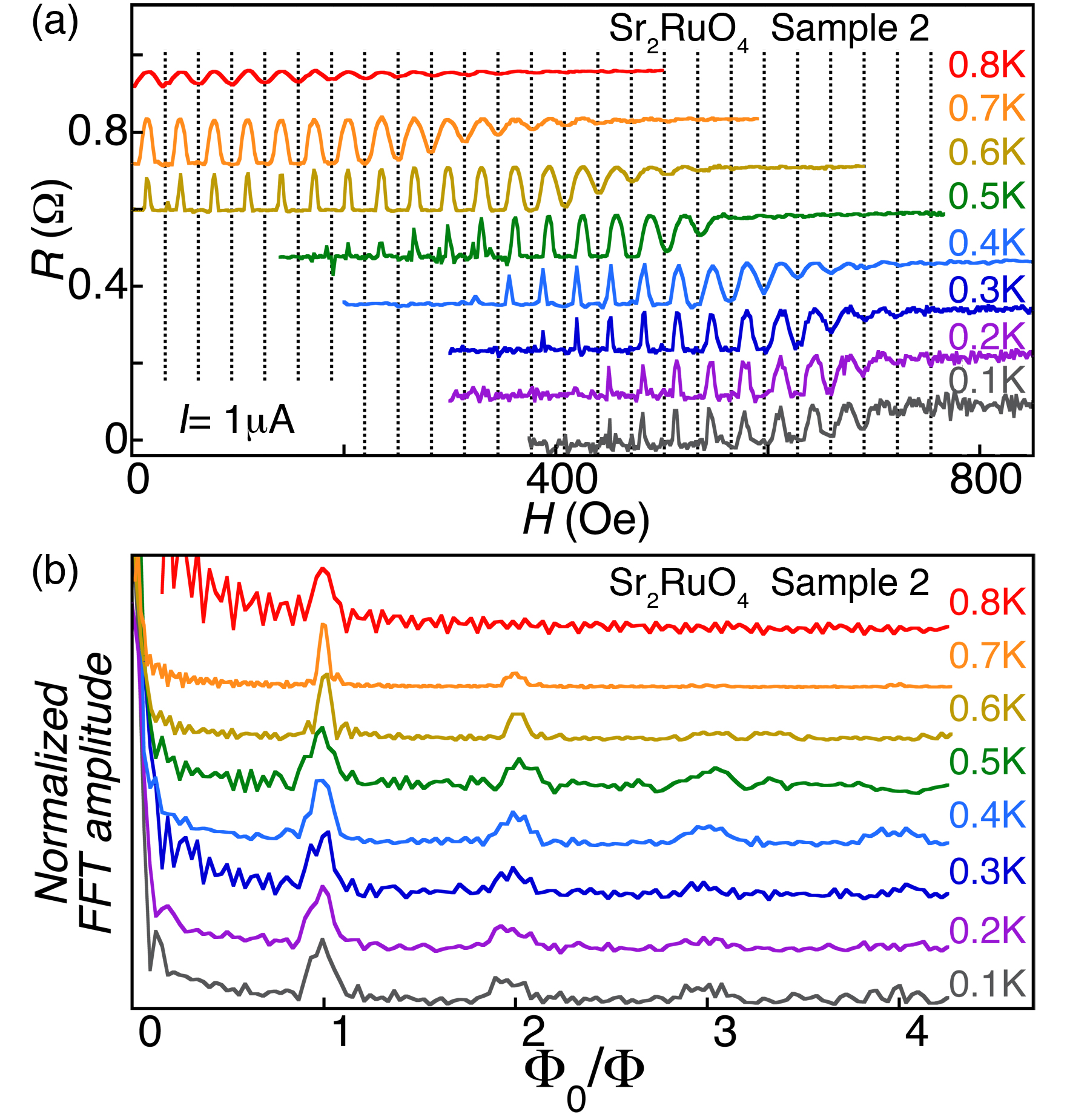}
\caption{\label{fig:epsart} (Color online) (a) Magnetic field dependence of sample resistance $R(H)$ for Sample 2 at various temperatures as indicated. The spacing of the dashed lines is $\Delta H$ = 31.4 Oe, the oscillation period corresponding to $\Phi_0$. (b) Normalized Fourier transform amplitude of the $R(H)$ oscillations as a function of inverse magnetic flux at various temperatures as indicated. The horizontal axis values were obtained by multiplying inverse field $1/H$ by $\Delta H$ = 31.4 Oe. The amplitude of the dominant peak was normalized to 1 for each temperature. Curves are shifted vertically for clarity.}
\end{figure}

\begin{figure}[t!]
\includegraphics[width=3.36in]{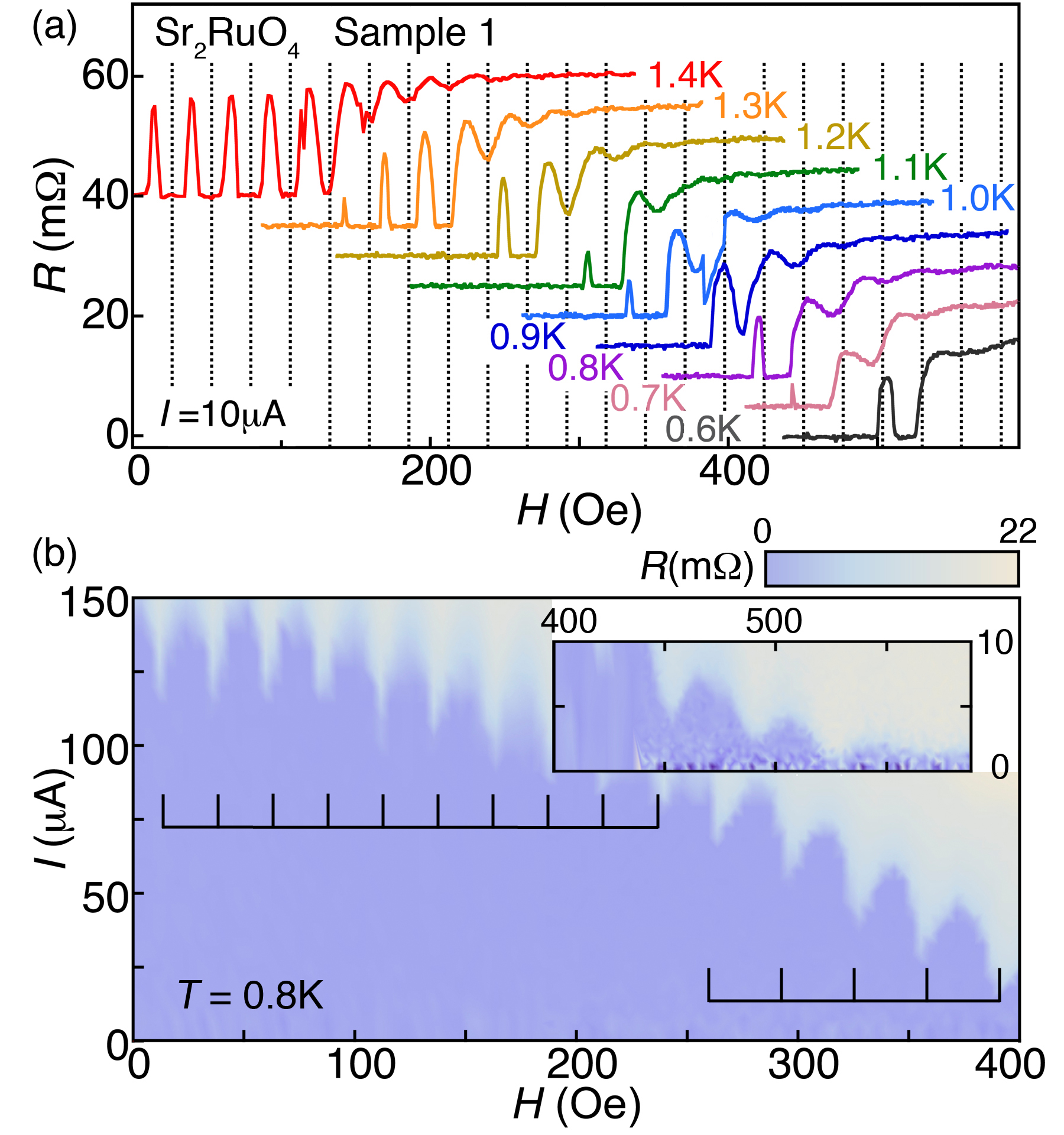}
\caption{\label{fig:epsart} (Color online)  (a) Magnetic field dependence of sample resistance $R(H)$ for Sample 1 at various temperatures as indicated. The spacing of the dashed lines is $\Delta H$ =26.3 Oe, the oscillation period corresponding to $\Phi_0$. (b) Color density plot of the sample resistance as a function of applied current $(I)$ and magnetic field $(H)$ for Sample 1 at 0.8 K showing two distinct periodicities. The spacing between the solid bars is 25.3 Oe at low fields and 32.4 Oe at high fields, respectively. The inset shows the resistance color map in the same field range as resistance oscillations appear in (a). }
\end{figure}

We performed a Fast Fourier Transform (FFT) of our data to seek out evidence for a $\Phi_0$/2 oscillation period. As shown in Fig.~3(b), the frequency corresponding to $\Phi_0$ oscillations clearly dominates. The much smaller peak for $\Phi_0$/2 oscillations and other peaks are most likely higher harmonics of the $\Phi_0$ oscillation. On the other hand, given that the amplitude of the $\Phi_0$/2 oscillation is not known, in principle the existence of a small $\Phi_0$/2 oscillation cannot be excluded. The observation of the splitting of resistance peaks as observed previously in a different material system~\cite{Zadorozhny2001} could avoid this uncertainty. However, no such splitting was confirmed in our data. The absence of a splitting in resistance peaks suggests that the application of an in-plane magnetic field used in the previous cantilever magnetometry experiment~\cite{Jang2011,Vakaryuk2011} may indeed be crucial in stabilizing $\Phi_0$/2 states.

In Sr$_2$RuO$_4$ rings with a thick wall, a sudden increase in the period of the resistance oscillations was observed [Fig.~4(a)]. The estimated $\Phi_0$ oscillations of this sample from Eq.~(1) result in a period of $26 \pm 2$ Oe. In low magnetic fields, resistance oscillations with a period $\Delta H \approx$ 26.3 Oe, as indicated by the dashed lines in Fig. 4(a), were clearly observed when $T$ was close to $T_c$. For the resistance oscillations at larger fields, it is necessary to measure at lower temperatures in order to place the the resistive transition of the sample at the larger fields. A different period of 32 Oe in resistance oscillations were found. 

The existence of two oscillation periods is confirmed in the measurements of voltage-current $(V-I)$ curves that yield the superconducting phase boundary at temperatures far below $T_c$. We carried out these measurements at closely spaced magnetic fields ($H$) for $T=$0.8 K [Fig. 4(b)]. Here the color code represents the value of sample resistance. The oscillating phase boundary extended over the whole field range down to zero field. Two distinct periodicities were observed, $\Delta H \approx$ 25.3 Oe at low fields and $\Delta H \approx$ 32.4 Oe at high fields. While the former is consistent with the conventional $\Phi_0$ oscillations (see above), the latter is larger than that expected from $\Phi_0$ resistance oscillations. The change in the periodicity of the critical current oscillations, at a field around 250 Oe, is rather sharp. A similarly sharp change in the periodicity of the resistance oscillations has also been observed in a different sample (Sample 3) at around 100 Oe. The periods are about 25 Oe and 36 Oe, respectively. This abrupt increase in the oscillation period can not be explained by the phase diagram for a conventional Type II superconducting rings \cite{Baelus2000}, nor other factors such as the demagnetization effect at the edge of mesoscopic samples. 

We propose the following picture to explain the observed change in periodicity. In an applied magnetic field, previous studies of the few-vortex states of mesoscopic Type II superconductors suggest that the free energy of the sample will be minimized if the ring is decorated by vortices~\cite{Geim1997}. In small fields, these vortices will be driven in and out of the ring because of the weak pinning potentials in crystalline Sr$_2$RuO$_4$ and the lack of sample space to form a square vortex lattice expected for this field regime~\cite{Curran2011}. For the two thick-wall rings, on which two different periods of resistance oscillations were found, we note that the wall thickness, ($w\sim 4.8\xi(0)$) and ($w\sim 5.5\xi(0)$), respectively, would allow two vortices along the radial direction of the ring (the nominal size of the normal core of a vortex is $2\xi(0)$). We speculate that a vortex ``lock-in" occurs when two loops of vortices are allowed. This ``lock-in" of vortices could be facilitated by collective vortex pinning that is effective when a vortex lattice is formed~\cite{Blatter1994}. Because of the normal cores of the vortices, the ``lock-in" of the vortices reduces the effective wall thickness of the ring and therefore increases the oscillation period, $\Delta H$, based on Eq. (1). The abrupt change in periodicity therefore indicates the ``lock-in" of the vortex lattice occurs in a narrow regime of applied field. Interestingly, the ``lock-in" fields observed for our two samples are consistent with the idea that a large field is needed to form a vortex lattice for a ring of a thinner wall. Furthermore, it is also natural for a thin-wall ring such as Sample 2 to feature a single periodicity in quantum oscillations because only a single vortex is allowed along a radial direction up to the upper critical field of Sr$_2$RuO$_4$. On the other hand, more detailed theoretical and experimental studies are needed to verify this vortex lattice ``lock-in" picture.  

In summary, we carried out the first magnetoresistance measurements of mesoscopic superconducting rings of Sr$_2$RuO$_4$. Thin-wall rings were found to exhibit quantum oscillations of a single period of $\Phi_0$. Large amplitude as well as large number of the quantum oscillations were found, which are explained in a vortex flow rather than conventional LP effect picture. For rings with a thick wall, two distinct periods of quantum oscillations were found in high and low-magnetic filed regimes, respectively, which are attributed to the ``lock-in" of a vortex lattice in the ring. In either type of the samples, however, no evidence for $\Phi_0$/2 resistance oscillations was identified, suggesting that an in-plane field may indeed be crucial for the observation of the $\Phi_0$/2 flux states. 

We acknowledge useful discussions with M. Sigrist, C.-C. Tsuei, Y. Maeno, S.-K. Yip, V. Vakaryuk, Z. Wang, and C. Nayak and technical assistance on device characterization from T. Clark and J. Kulik. Work at Penn State is supported by DOE under Grant No. DE-FG02-04ER46159 and at Tulane by NSF under grant DMR-1205469 and DOD-ARO under Grant No. W911NF0910530. Nanofabrication and structural characterization work was done at Penn State with NSF support under Grant No. DMR 0908700 and the Penn State MRI Nanofabrication Lab under NSF Cooperative Agreement 0335765, NNIN with Cornell University, and at Florida State University supported by the Florida State University Research Foundation, NSF-DMR-0654118, and the State of Florida.

\nocite{*}

\bibliography{refs.bib}

\end{document}